\documentclass[aps,prd,superscriptaddress,amsfonts,amssymb,
amsmath,showpacs,floatfix,preprintnumbers,twocolumn,nofootinbib]{revtex4-2}

\usepackage{dcolumn}
\usepackage{bm}
\usepackage{multirow}
\usepackage{verbatim}
\usepackage{graphicx}
\usepackage[normalem]{ulem}
\usepackage{color}
\usepackage{bbm}
\usepackage{slashed}
\usepackage{braket}
\usepackage{fancyhdr}
\usepackage{subcaption}
\usepackage{float}
\usepackage{caption}
\usepackage{makecell} 

\newcommand{\be}{\begin{equation}}
\newcommand{\ee}{\end{equation}}

\begin{document}
\rightline{TTK-24-44}

\title{Efficient State Preparation for the Schwinger Model with a Theta Term}



\author{Alexei Bazavov}
\affiliation{Department of Computational Mathematics, Science, Michigan State University, East Lansing, Michigan
48824, USA}
\affiliation{Department of Physics and Astronomy, Michigan State University, East Lansing, Michigan 48824, USA}

\author{Brandon Henke}
\affiliation{Department of Physics and Astronomy, Michigan State University, East Lansing, Michigan 48824, USA}

\author{Leon Hostetler}
\affiliation{Department of Physics, Indiana University, Bloomington, Indiana 47405, USA}

\author{Dean Lee}
\affiliation{Facility for Rare Isotope Beams and Department of Physics and Astronomy, Michigan State University, East Lansing, Michigan 48824, USA}

\author{Huey-Wen Lin}
\affiliation{Department of Physics and Astronomy, Michigan State University, East Lansing, Michigan 48824, USA}

\author{Giovanni Pederiva}
\email[e-mail: ]{g.pederiva@fz-juelich.de}
\affiliation{J\"ulich Supercomputing Center, Forschungszentrum J\"ulich, Wilhelm-Johnen-Stra{\ss}e, 54245 J\"ulich, German}
\affiliation{Center for Advanced Simulation and Analytics (CASA), Forschungszentrum J\"ulich, Wilhelm-Johnen-Stra{\ss}e, 52425 J\"ulich, Germany}

\author{Andrea Shindler}
\email[e-mail: ]{shindler@physik.rwth-aachen.de}

\affiliation{Institute for Theoretical Particle Physics and Cosmology, TTK, 
RWTH Aachen University, Sommerfeldstr. 16, Aachen, 52074, Germany}

\affiliation{Nuclear Science Division, Lawrence Berkeley National Laboratory, 
Berkeley, CA 94720, USA}%

\affiliation{Department of Physics, University of California, Berkeley, CA 94720, USA}%

\begin{abstract}
	We present a comparison of different quantum state preparation algorithms and
	their overall efficiency for the Schwinger model with a theta term. While adiabatic state preparation (ASP) is proved to be effective, in practice it leads to large CNOT gate counts to prepare the ground state.
	The quantum approximate optimization algorithm (QAOA) provides excellent results while keeping the CNOT counts small by design, at the cost of an expensive classical minimization process. We introduce a ``blocked'' modification of the Schwinger Hamiltonian to be used in the QAOA that further decreases the length of the algorithms as the size of the problem is increased. The rodeo algorithm (RA) provides a powerful tool to efficiently prepare any eigenstate of the Hamiltonian, as long as its overlap with the initial guess is large enough.
	We obtain the best results when combining the blocked QAOA ansatz and the RA, as this provides an excellent initial state with a relatively short algorithm without the need to perform any classical steps for large problem sizes.

\end{abstract}

\maketitle

\section{Introduction}
\label{sec:intro}

Despite the enormous success of the Standard Model (SM) of particle physics, many unanswered
questions remain, such as what constitutes dark matter (and dark energy),
or the observed amount of matter-antimatter asymmetry of the Universe.
It was established many years ago, thanks to the work of Sakharov~\cite{Sakharov:1967dj},
that the fundamental interaction and mechanism at the origin of the baryon asymmetry in the Universe
must satisfy three conditions, known as the Sakharov conditions.
An important aspect of these conditions is that the fundamental interaction should
break CP symmetry. While the SM contains, due to quark-flavor mixing, a direct source
of CP violation, the strength of this violation is not sufficient to explain the
observed asymmetry~\cite{Gavela:1993ts,Huet:1994jb}.
The search for physics beyond the Standard Model (BSM) thus includes a search for
new sources of CP violation. A prototypical example of a CP-violating interaction
that has several phenomenological implications (e.g.~axions, electric dipole moments, 
the baryon asymmetry of the universe)
is the $\theta$ term of the strong interactions described by quantum chromodynamics (QCD).

QCD is a theory asymptotically free at short distances, where it is amenable to a perturbative treatment,
but at low energy becomes nonperturbative, and to this day the only way to perform calculations
of QCD at low energy with systematically improvable uncertainties
is to regulate the theory on a four-dimensional lattice (lattice QCD) and solve it numerically.
Lattice QCD (LQCD) is a mature field with a well-defined selection of problems that it can address,
but also with clear obstacles that nowadays seem insurmountable. One of these
difficulties is the impossibility to solve theories that possess a complex Euclidean action,
because it prevents the use of stochastic methods to properly sample the field space.
A classic example of complex action in Euclidean space is QCD with a $\theta$ term.
To circumvent this problem, it is common to expand the theory in powers of the
parameter $\theta$, which is known to be smaller than about 
$10^{-10}$~\cite{Abel:2020gbr,Dragos:2019oxn},
and treat the CP-violating interaction as a perturbation of QCD.\ 
While calculations of the neutron electric dipole moment have been done this way
~\cite{Dragos:2019oxn,Alexandrou:2020mds,Bhattacharya:2021lol,Liang:2023jfj}, 
confirming the smallness of the $\theta$ parameter, 
ideally we would like to
have a computational framework that is able to deal with a generic complex action,
or even better that does not need the theory to be rotated to Euclidean space,
rather allowing real-time simulations of the theory prepared in Minkowski space.
The impossibility of simulating complex actions is just one of the aspects of what nowadays
goes under the very general definition of ``sign problem''.\

A solution, at least in theory, is represented by the Hamiltonian formalism, which deals
directly with real-time systems, but it cannot be simulated using state-of-the-art supercomputers,
due to the large number of degrees of freedom stemming from the regulated infinite-dimensional
Hilbert space. The simulation of a theory using the Hamiltonian formalism seems to be a problem
perfectly suited for a quantum computer, and even though the number of qubits currently available
is still far too low to directly simulate QCD in $3+1$ dimensions,
it is possible to use lower-dimensional field theories that share certain properties with QCD
to test new ideas and algorithms.
It is particularly important to understand the scaling of the algorithms with increasing
size of the system and towards the continuum limit.

In this work we study quantum electrodynamics (QED) in $1+1$ dimensions,
also known as the Schwinger model~\cite{Schwinger:1962tp},
with the addition of a $\theta$ term.
The Schwinger model shares many interesting properties with QCD,
such as confinement of fermions and the spontaneous breaking of the $\mathrm{U}(1)$ symmetry
with a corresponding chiral condensate, so it is an ideal toy model to test
an algorithmic or computational paradigm.

The field of quantum computing for field theories, like the Schwinger model~\cite{Chakraborty:2020uhf,Farrell:2024fit,Desaules:2022ibp,Bauer:2023qgm,Farrell:2023fgd,Pomarico:2023png,Florio:2023dke,Xie:2022jgj,Nguyen:2021hyk,deJong:2021wsd,Honda:2021ovk,Honda:2021aum,Banuls:2019bmf,Kokail:2018eiw,Klco:2018kyo,Lu:2018pjk,Martinez:2016yna,Kuhn:2014rha,Hauke:2013jga}, is making significant advances. A possible approach is to rely on discretizing the spatial coordinates, allowing the form of the discretized Hamiltonians of the system to be determined using standard techniques currently employed in lattice QCD. We have used the Schwinger model with a $\theta$ term to test several quantum algorithms for quantum state preparation with particular emphasis on the scaling
with the size of the problem and the number of qubits.
In Sec.~\ref{sec:model} we introduce the discretization of the Schwinger model on a quantum system,
while in Sec.~\ref{sec:asp} we study the first algorithm, the adiabatic state evolution.
In Sec.~\ref{sec:qaoa} we study the quantum approximate optimization algorithm (QAOA)
and in Sec.~\ref{sec:rodeo} the newly proposed Rodeo algorithm~\cite{Choi:2020pdg}.
In Sec.~\ref{sec:conclusion} we present our summary and conclusions.


\section{Discretization of the Schwinger Model on a Quantum System}
\label{sec:model}
\subsection{The Continuum Schwinger Model}

The Schwinger model is the theory of QED in $1+1$ dimensions. In the presence of a topological $\theta$ term its Lagrangian is typically written as
\begin{equation}
\mathcal{L} = -\frac{1}{4}F_{\mu\nu}F^{\mu\nu} + \frac{g\theta}{4\pi}\epsilon_{\mu\nu}F^{\mu\nu} + i\bar{\psi}\gamma^\mu(\partial_{\mu} + igA_{\mu})\psi - m\bar{\psi}\psi,
\end{equation}
where $F_{\mu\nu} = \partial_{\mu}A_{\nu} - \partial_{\nu}A_{\mu}$ is the field tensor, the $A_{\mu}$ are $\mathrm{U}(1)$ gauge fields, and $\epsilon_{\mu\nu}$ is the totally antisymmetric tensor. In $1+1$ dimensions, the gamma matrices are $\gamma^0=\sigma^z$, $\gamma^1=i\sigma^y$, and $\gamma^5=\gamma^0\gamma^1$.

The theory has three parameters: the gauge coupling $g$, the fermion mass $m$, and the $\theta$ angle. Following prior works~\cite{Chakraborty:2020uhf}, we perform a chiral transformation of the fields $\psi \rightarrow e^{i\frac{\theta}{2}\gamma_5} \psi$, $\bar{\psi} \rightarrow \bar{\psi} e^{i\frac{\theta}{2}\gamma_5}$ and the path integral measure~\cite{Fujikawa:1979ay,Roskies:1981} to arrive at an equivalent Lagrangian
\begin{equation}
\mathcal{L} = -\frac{1}{4}F_{\mu\nu}F^{\mu\nu} + i\bar{\psi}\gamma^\mu(\partial_{\mu} + igA_{\mu})\psi - m\bar{\psi}e^{i\theta\gamma_5}\psi.
\end{equation}
We choose the temporal gauge $A_0 = 0$, and then a standard Legendre transform yields the Hamiltonian
\begin{equation}
\label{eq:cont_hamiltonian}
H = \int dx \left[ -i\bar{\psi}\gamma^1 (\partial_1 + igA_1)\psi + m\bar{\psi}e^{i \theta \gamma_5}\psi + \frac{1}{2} E^2 \right],
\end{equation}
where in one spatial dimension, the electric field $E=F^{10}=-\dot{A}^1$ has only one component, and there is no magnetic field. To satisfy gauge invariance in the temporal gauge, additional local constraints that govern the interaction between matter and gauge fields must be imposed. These constraints are provided by the Gauss law $\partial_1 E(x) = g \bar\psi(x)\gamma^0 \psi(x)$.

\subsection{Discretization}

To simulate the Schwinger model on a quantum device, we need a discretized formulation of the Hamiltonian Eq.~(\ref{eq:cont_hamiltonian}). We use a 1-dimensional spatial lattice with $N$ sites and lattice spacing $a$. Time is kept continuous. Following prior works~\cite{martinez2016,muschik2017u,Chakraborty:2020uhf} we use Kogut-Susskind staggered fermions~\cite{kogut1975hamiltonian,susskind1977lattice} to arrive at the Hamiltonian
\begin{align}
\nonumber
H &= -i\sum_{n=1}^{N-1}\left(\frac{1}{2a} - (-1)^n\frac{m}{2}\sin\theta \right) \left[\chi_n^\dagger e^{i\phi_n}\chi_{n+1} - \text{h.c.} \right] \\
 & \qquad + m\cos\theta\sum_{n=1}^N {(-1)}^n\chi_n^\dagger \chi_{n} + \frac{g^2a}{2} \sum_{n=1}^{N-1} L^2_n.
\end{align}
The gauge operators have been rescaled as $A^1(x_n) \rightarrow -\phi_n/(ag)$ and $E(x_n) \rightarrow gL_n$, where $\phi_n$ lives on site $n$, and $L_n$ lives on the link between sites $n$ and $n+1$. The Dirac fermion $\psi(x) = (\psi_u(x),\psi_d(x))^T$, which is a 2-component spinor in $1+1$ dimensions, has been mapped to a pair of 1-component fermions $\chi_n$ living on neighboring sites such that $\chi_n = \sqrt{a}\psi_u(x_n)$ for even $n$ and $\chi_n = \sqrt{a}\psi_d(x_n)$ for odd $n$. In this formulation the Gauss law becomes~\cite{muschik2017u}
\begin{equation}
L_n - L_{n-1} = \chi_n^\dagger \chi_{n} - \frac{1-(-1)^n}{2} .
\end{equation}
For a given matter configuration, the gauge fields are now completely determined\footnote{This is true when using open boundary conditions as in this work.}.

\subsection{Spin Hamiltonian}

The $\chi$ field can be transformed into a qubit formulation
using the Jordan-Wigner transformation~\cite{wigner1928paulische}
that transforms the fermionic variables into spin variables
\begin{equation}
	\chi_n = \left(\prod_{l<n} -i Z_l\right) \frac{X_n - i Y_n}{2},
\end{equation}
where the spin variables are the Pauli matrices located at each lattice
point, $X_i=\sigma_i^x$, $Y_i=\sigma_i^y$, $Z_i=\sigma_i^z$.
Using open boundary conditions, i.e.~fixing the conjugate momentum $L$ at the
boundary, and solving the Gauss law, one obtains
\begin{equation}
	L_n = L_0 + \frac{1}{2}\sum_{l=1}^n\left( Z_l + (-1)^l \right)\,,
\end{equation}
where the value of $L_0$ specifies the boundary conditions.
Removing $L_0$ is equivalent to shifting the $\theta$ angle by
$2\pi L_0$~\cite{coleman:1976uz},
thus, we can safely set $L_0 = 0$.
The $\phi$ phases can be absorbed into the fields by a gauge transformation
$\chi_n \rightarrow \prod_{l<n}[e^{-i\phi_n}]\chi_n$.

The final Hamiltonian, omitting constant terms, can be decomposed as
$H = H_{ZZ} + H_{\pm} + H_{Z}$, where
\begin{align}
	\label{eq:hamiltonian}
	\nonumber
	H_{ZZ}  & = \frac{J}{2}\sum_{n=2}^{N-1}\sum_{1\leq k < l \leq n} Z_k Z_l \\
	H_{\pm} & = \frac{1}{2}\sum_{n=1}^{N-1}
	\left(w - (-1)^n\frac{m}{2}\sin\theta \right)
	\left[X_n X_{n+1} + Y_n Y_{n+1} \right]                                  \\
	\nonumber
	H_{Z}   & = \frac{m\cos\theta}{2} \sum_{n=1}^N (-1)^n Z_n
	-\frac{J}{2}\sum_{n=1}^{N-1} (n ~\text{mod} ~2) \sum_{l=1}^n Z_l\,,
\end{align}
where, using the same notation as Ref.~\cite{Chakraborty:2020uhf}, we denote
the relevant couplings for the adiabatic evolution as $w=1/(2a)$ and $J=g^2a/2$.

\section{Adiabatic State Preparation}
\label{sec:asp}
Adiabatic state preparation (ASP) is a common reference
when dealing with state-preparation algorithms,
because of its relative simplicity and robustness.
The algorithm allows one to initialize a set of qubits
to the ground state of a chosen ``target'' Hamiltonian
without the need for any additional qubits.
What is needed instead is to find a ``simple'' Hamiltonian
that possesses a readily achievable ground state.
The simple Hamiltonian should be related to the full target
one in such a way that a set of coefficients can be used to interpolate between them.

ASP works by applying a set of time evolution
operators each with a different Hamiltonian, which come from a
discretization of the interpolation between the simple and the target Hamiltonians.
This allows the state to evolve towards the desired target ground state
by always remaining close to the ground state of the interpolating Hamiltonians
at all time steps.

In the Schwinger model, we select the initial Hamiltonian as follows:
\begin{equation}
    H_0 = H_{ZZ} + H_Z|_{m\rightarrow m_0, \theta \rightarrow 0}\,,
\end{equation}
whose ground state is a state of alternating spins.
The parameters of the Hamiltonian, $w=1/(2a)$, $\theta$ and $m$,
are rewritten to be adiabatic dependent on an additional
variable called ``adiabatic time'' $t$:
\begin{equation}
        w\rightarrow \frac{t_i}{T} w\quad\theta\rightarrow \frac{t_i}{T} \theta\quad
    m\rightarrow \left(1 - \frac{t_i}{T} \right)m_0 + \frac{t_i}{T} m\,,
    \label{eq:linear}
\end{equation}
for a set of $i=1,\ldots, M$ values of the adiabatic time $t_i = t_{i-1} + \delta t_i$
with time intervals $\delta t_i$. The end time of the adiabatic evolution is
denoted by $T = \sum_{i=1}^{M} \delta t_i$.

The simplest discretization is a linear one,
corresponding to taking all $\delta t_i$ to be the same, $\delta t_i = T/M$,
but other nonuniform values for $\delta t_i$ can be chosen.

To prepare the desired quantum state, we define an adiabatic time-evolution operator $U(t, t+\delta t) = e^{-iH_A(t)\delta t}$,
where $H_A(t)$ is the interpolating adiabatic Hamiltonian at adiabatic time $t$.
We distinguish between two types of linear discretizations
of the adiabatic evolution, labeled ``L1'' and ``L2'' according to
the order of the Trotter product formula used to
evaluate the evolution operator $U(t, t+\delta t)$.

We also consider other discretizations in this paper, given by
\begin{equation}
\label{eq:sin2}
\delta t_i = 2 \frac{T}{M} \sin^2\left(\pi \frac{i}{M}\right)\,,
\end{equation}
\begin{equation}
\label{eq:cos2}
\delta t_i = 2 \frac{T}{M}\cos^2\left(\pi \frac{i}{2M}\right)\,,
\end{equation}
and
which we denote respectively by ``S1'' (or ``S2'') and ``C1'' (or ``C2'')
depending on whether we use a first-order or a second-order
Trotter product formula for the adiabatic time evolution.

These choices are made in order to achieve discretizations
that are either denser or sparser at different stages of the adiabatic evolution.
The S1, S2 discretizations are denser at the beginning and the end of the
adiabatic evolution, while the C1, C2 choices are sparser at the beginning
of the adiabatic evolution and denser at the end.

It is common practice, in the field of quantum computing, to count the number of two-qubit gates, typically CNOT gates, as a quantity that measures the length of the algorithm as they have a smaller fidelity on current quantum hardware and hence dominate the noise. For the Hamiltonian we are considering, the number of CNOT gates required for a single first-order Trotter step are given by $4(N-1) + (N-1)(N-2)$, where $N$ is the number of sites of the lattice. The first term comes from the hopping term $H{\pm}$; the second comes from the nonlocal term $H_{ZZ}$ and crucially grows quadratically in $N$. The mapping of the terms of the Hamiltonian to quantum gates is explained in~\cite{Chakraborty:2020uhf}.

For reference, in Tab.~\ref{tab:cnot}, we report the average CNOT gate count per qubit for each step of the adiabatic time 
evolution for system with $N=4,8,12$ and both 
first- and second-order Trotter discretizations, which differ by a simple factor of two. The quadratic scaling coming from $H_{ZZ}$ dominates already for small values of $N$, making the average number of gates grow linearly with the size of the problem.
\begin{table}[h]
    \begin{center}
        \begin{tabular}{ccc}
            $N$  & 1st Order & 2nd Order \\\hline 
            $4$  & $4.5$     & $9$       \\
            $8$  & $8.8$     & $17.5$    \\
            $12$ & $12.8$    & $25.7$     
        \end{tabular}
        \caption{Average number of CNOT gates per qubit for each step of the adiabatic time.
        $N$ is the number of points in the spatial direction and the order refers to trotterization adopted.}
        \label{tab:cnot}
    \end{center}
\end{table}

\subsection{Numerical Results}

We investigate the quantum algorithm for ASP
on a classical computer, exploring how the choice of discretization and total
evolution time $T$ influence the quality of the prepared ground state.
To assess our algorithms, we consider the energy and overlap with the
exact ground state obtained from diagonalization. We focus in particular
on small values of $M$ and $T$ for two systems with $N=4,8$, which lead to
short overall algorithms.

We use the Qiskit software package~\cite{qiskit} for our numerical simulations. 
This choice allows to use a gate-based representation of the algorithms and to also 
access the state vector during the simulations.
There are two parameters that significantly impact the final approximation.
The first is the ratio $\delta t / T$, which determines the speed
of the system's evolution between the initial and the target Hamiltonian.
A large ratio reduces the number of steps required,
but at the cost of lower precision in the intermediate states,
as they are further away from their eigenstates.
A small ratio, on the other hand, ensures the state remains
close to the intermediate ground states during the evolution,
but requires more steps to achieve convergence.

The second parameter is $\delta t$, which affects the ASP algorithm
by influencing the accuracy of the time evolution operators
used to approximate the Hamiltonian with the Trotter product formula.
This error can be reduced with higher-order schemes,
but this comes at the cost of longer algorithms per time step.

\begin{figure}[!hbpt]
    \centering
    \includegraphics[width=1\linewidth]{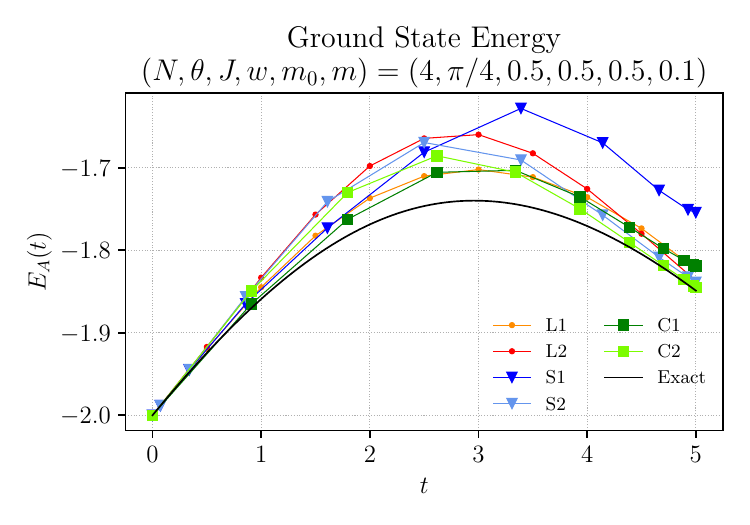}
    \includegraphics[width=1\linewidth]{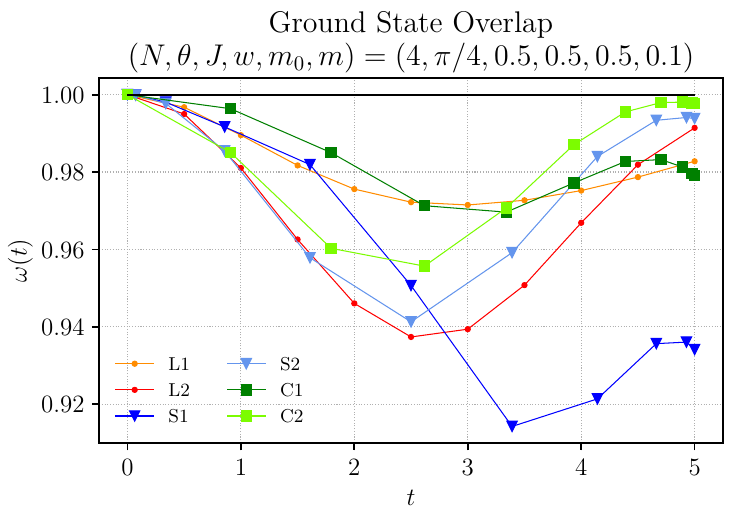}
    \caption{\footnotesize
    Ground-state energy $E_A(t)$ (top) and overlap with the true ground state,  $\omega(t)$ (bottom),
    versus adiabatic time $t$ for the
    Schwinger model with a theta term.
    The exact results, represented by the solid line,
    are obtained through exact diagonalization of the adiabatic Hamiltonian
    at each value of $t$.
    Various discretizations are employed,
    as explained in the main text.}
    \label{fig:asp_energy}
\end{figure}

In Fig.~\ref{fig:asp_energy}, we present the time evolution
of the ground-state energy, $E_A(t)$,
using different discretizations with $T=5$ and $M=10$.
The exact values are computed by direct diagonalization of the intermediate Hamiltonian
at every time step. As shown, all the discretizations exhibit good agreement with
the exact results even with as few as $M=10$ steps.
However, we observe that the C2 and C1
discretizations yield better agreement, indicating that it may be beneficial to
perform more precise steps towards the end of the evolution, where the Hamiltonian
is closer to the target one.

The overlap $\omega(t)$ between the evolved
state $\ket{\psi_A(t)}$ and the exact ground state
$\ket{\psi_0(t)}$ of the adiabatic Hamiltonian
obtained via full diagonalization at each time $t$, is
$\omega(t)=|\braket{\psi_A(t)|\psi_0(t)}|^2$.
The plot reveals that $\omega$ is always within $10\%$
of the exact ground state,
indicating that the evolution successfully
keeps the state close to the intermediate ground
state throughout the process.
Notably, we find that discretization C1 offers
the most precision throughout the adiabatic
evolution but loses precision towards the end,
whereas C2 yields the best approximation at the end of the evolution.

\begin{figure}[!htpb]
    \centering
    \includegraphics[width=1\linewidth]{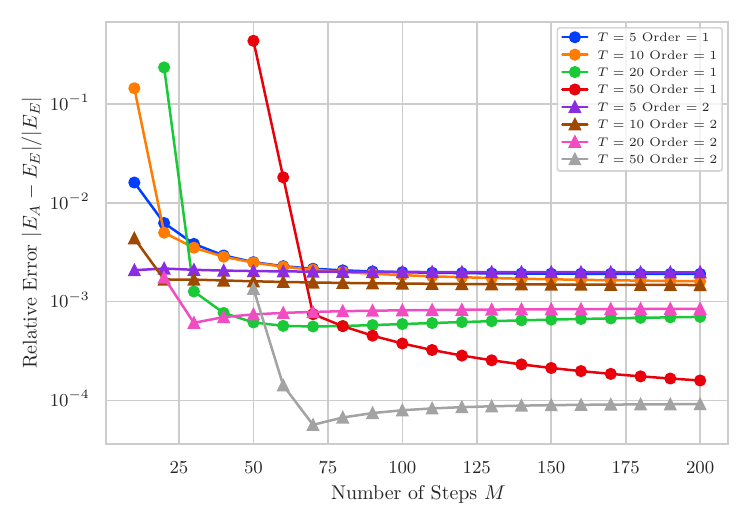}
    \includegraphics[width=1\linewidth]{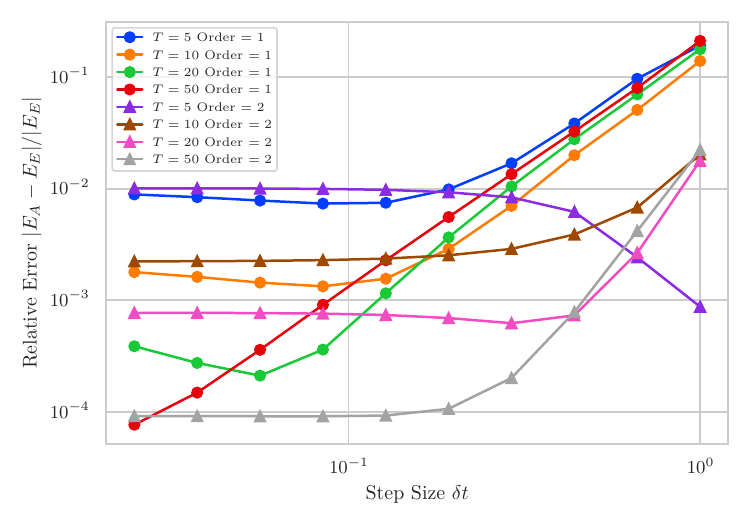}
    \caption{\footnotesize
    Relative error of the ground state energy for the same simulation parameters as Fig.~\ref{fig:asp_energy} for the L1 and L2 discretizations. The top figure shows the error as a function of the total number of steps ($T/\delta T$) for a fixed set of total times $T$. The lower panel shows the error just as a function of $\delta T$ for the same set of values for $T$.}
    \label{fig:asp_convergence}
\end{figure}

We investigate the dependence of the total adiabatic time evolution $T$
on the variation of the number of steps $M$.
The top panel of Fig.~\ref{fig:asp_convergence} displays the relative error
for the ground state energy $E_A(T)$ obtained at the end of the
adiabatic evolution as a function of $M$.
The different colors correspond to distinct adiabatic time evolutions $T$,
with linear discretization orders of L1 or L2.
As expected, a larger $T$ results in smaller errors.
However, we also observe that there is a threshold value of $M$ for a
given $T$ at which the relative error is minimized and after which increasing the number of steps indefinitely
does not further reduce the relative error. This is fixed by $T$ and is unaffected by the order of the trotterization.

In the bottom panel of Fig.~\ref{fig:asp_convergence} the difference between first-order and second-order trotterization is apparent, as the two classes have different slopes. This indicates that using second-order trotterization allows smaller values of $\delta t$, as expected, as long as the total adiabatic time $T$ is large enough for the algorithm to be in the scaling regime.

Overall, ASP is a reliable method for preparing the ground state of our 
Hamiltonian, however, the required algorithm is limited by the bounds
imposed by the adiabatic theorem, which then requires longer evolution
times and hence more steps and longer quantum algorithms.

\subsection{Noise Model for Adiabatic State Preparation}
\label{subsec:asp_noise}

\begin{figure}[h!]
    \centering
    \includegraphics[width=1\linewidth]{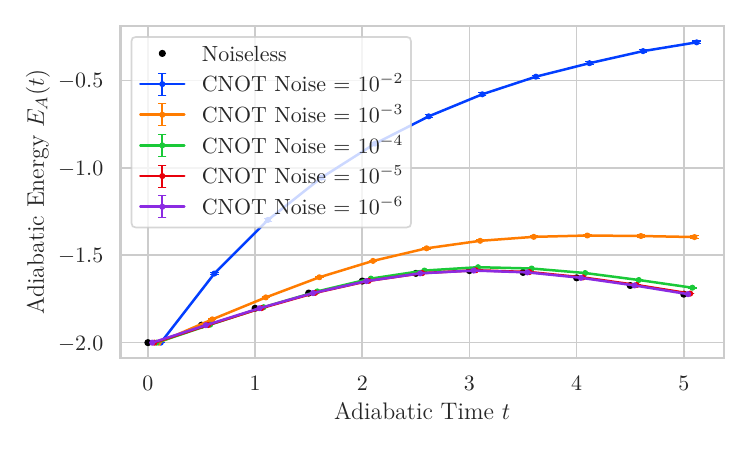}
    \caption{\footnotesize
    Ground state energy from ASP with L2 discretization with varying CNOT
    error rate. The parameters chosen are the same as in Fig.~\ref{fig:asp_energy}
    with $T=5$ and $M=10$.}
    \label{fig:asp_noise}
\end{figure}

One major challenge of ASP is the length of the algorithm resulting
from the adiabatic time evolution discretization.
On noisy intermediate-scale quantum (NISQ) machines, such algorithms can become prohibitively
expensive due to the presence of many CNOT gates with relatively low precision.
To investigate the impact of CNOT gate errors on ASP,
we conducted a series of simulations using Qiskit (using
discretization L2)
where we varied the error rate of the CNOT gates.
Fig.~\ref{fig:asp_noise} shows the results of this study for
$T=5$ and $M=10$ where
we varied the error rate of CNOT gates from $10^{-2}$ to $0$, i.e~ noiseless.
The goal is to determine what error rate would allow the ASP algorithm
to produce a good estimate of the ground state of the $N=4$ Schwinger model.
The plot compares the resulting ground-state energy to the exact results,
and it is clear that, at current error rates 
ranging from $10^{-2}$ to $10^{-3}$~\cite{Kandala2021}, 
the noise significantly hinders the algorithm's
ability to produce a high-quality state.

\begin{figure}[hbt!]
    \centering
    \includegraphics[width=1\linewidth]{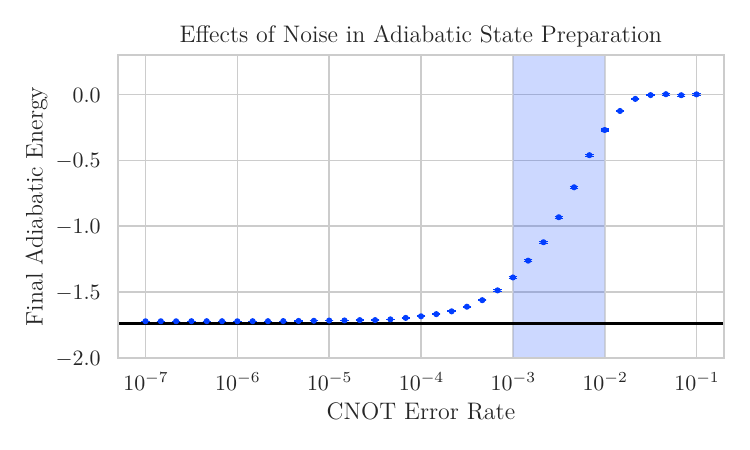}
    \caption{\footnotesize
    Energy of the ground state from ASP at the end of adiabatic
    evolution, $T=5$, with varying CNOT error rate.
    The band indicates state-of-the-art CNOT error gates~\cite{IBM2023}.}
    \label{fig:asp_noise_time}
\end{figure}

Fig.~\ref{fig:asp_noise_time} illustrates how the ground-state energy,
calculated with L2 discretization at the end of the
adiabatic evolution $T=5$ (with $M=10$)
depends on the error rate of the CNOT gates.
In the case of a large CNOT error rate,
the wave function tends towards a maximally disordered state,
resulting in an energy close to $0$.
The figure also demonstrates that extrapolating the error rates
from the current accessible ones (indicated by a band)
to regions where the results become independent on them is not currently possible.
It will be very difficult to perform an extrapolation
unless error rates improve by 1--2 orders of magnitude.

To summarize, the limits of current quantum hardware further limit the
applicability of ASP to our system, making a case for the need of shorter
and more efficient algorithms.



\section{Quantum Approximate Optimization Algorithm}
\label{sec:qaoa}
The quantum approximate optimization algorithm (QAOA)~\cite{farhi2019quantum} is a method that relies on the variational principle to solve optimization problems. State preparation can be cast to such a problem by giving a parametrized ansatz for the ground state wave function and then optimizing its parameters variationally. One of the most prominent benefits it offers is the relatively short length of the algorithms and the small number of parameters it requires.

Similar to ASP, the starting point is a trivially solvable Hamiltonian $H_0$ with eigenstate $\ket{\psi_0}$. The QAOA ansatz for the ground state of the target Hamiltonian is
\begin{equation}
    \label{eq:ansatz}
    \ket{\psi_M( \vec{\gamma}, \vec{\beta})} = \left(\prod_{k=0}^{M-1}e^{-i\beta_{M-k} H_0}e^{-i\gamma_{M-k} H}\right)\ket{\psi_0},
\end{equation}
where the $2M$ real coefficients $\vec{\beta}, \vec{\gamma}$ parametrize the wave function. From the variational principle, we know that given the parameters $\vec{\gamma}^*$ and $\vec{\beta}^*$, the expectation value of the Hamiltonian operator is
\begin{equation}
    \bra{\psi_M( \vec{\gamma}^*, \vec{\beta}^*)}H
    \ket{\psi_M( \vec{\gamma}^*, \vec{\beta}^*)} = E_0^V \geq E_0,
    \label{eq:qaoa}
\end{equation}
where $E_0$ is the true ground state of the system. This means that we can use a minimization algorithm; in our case, we used simulated annealing~\cite{kirkpatrick1983optimization}, because it is suitable for the multiple local minima of the problem. The minimization is performed classically and not on quantum hardware, though in the future a hybrid classical-quantum algorithm could be feasible as well.

As opposed to ASP, the length of the QAOA is chosen as a parameter instead of having to find the optimal number of Trotter steps. The precision of the results then depends on the quality of the optimal solution found by the minimizer, not on the length of the algorithm.

\subsection{QAOA Results}
We applied the QAOA to the case of the Schwinger model with the same parameters as Sec.~\ref{sec:asp}, so $(m,m_0,w,J,\theta) = (0,0.5,0.5,0.5,0)$ and $(1,0.5,0.5,0.5,\pi/4)$. Table~\ref{table:qaoa} displays the results for the energy expectation value and ground-state overlap for the ground state, obtained using the second-order Trotter formula, with two and three QAOA steps.

For the given ansatz, the total number of CNOT gates is given by $M(8(N-1) + (N-1)(N-2))$ for the first-order trotterization and twice as many for second-order. This is the same quadratic scaling as ASP, considering that the dominant term is given by $H_{ZZ}$. However, the total number of CNOT gates is much reduced because the number of steps $M$ is fixed beforehand to a very small number. We used $M=2,3$, which is orders of magnitude smaller than those used in ASP.\

\begin{table}[!hbtp]\footnotesize
    \centering
    \setlength\tabcolsep{5pt} 
    \begin{tabular}{ccccccc}
        \makecell{\\Method} & \makecell{\\$N$} & \makecell{\\$(\theta,m)$} & \makecell{\\$M$} & \makecell{CNOT/\\qubit} & \makecell{Rel.\\Err. $E_0$} & \makecell{GS\\Overlap} \\\hline 
        QAOA   & $4$ & $(0,0)$        & $2$ & $30$       & $0.0029$        & $0.9798$   \\
        QAOA   & $4$ & $(0,0)$        & $3$ & $45$       & $0.0012$        & $0.9928$   \\
        QAOA   & $4$ & $(\pi/4,0.1)$  & $2$ & $30$       & $0.0023$        & $0.9855$   \\
        QAOA   & $4$ & $(\pi/4,0.1)$  & $3$ & $45$       & $0.0021$        & $0.9871$   \\\hline
        QAOA   & $8$ & $(0,0)$        & $2$ & $49$       & $0.0021$        & $0.9679$   \\
        QAOA   & $8$ & $(0,0)$        & $3$ & $73.5$     & $0.0016$        & $0.9856$   \\
        QAOA   & $8$ & $(\pi/4,0.1)$  & $2$ & $49$       & $0.0077$        & $0.9732$   \\
        QAOA   & $8$ & $(\pi/4,0.1)$  & $3$ & $73.5$     & $0.0067$        & $0.9789$   \\
    \end{tabular}
    \caption{\footnotesize Results for the QAOA method for varying system sizes and parameters with first-order Trotter formula. Here $N$, $\theta$ and $m$ are the parameters of the simulation, and $M$ is the number of QAOA steps. The remaining columns give the CNOT gates per qubit, the relative error of the ground state energy measurement, and the ground state overlap.}~\label{table:qaoa}
\end{table}

The results show that QAOA is highly efficient in preparing the ground state and estimating its energy, using significantly fewer CNOT gates compared to ASP. Note that the values in Tab.~\ref{table:qaoa} refer to the total length of the algorithm, while in Tab.~\ref{tab:cnot} the count was per step.\ Specifically, for the $M=10$ case of ASP, the second-order approximation results in $90$ CNOT gates per qubit for $N=4$, leading to less accurate results when compared to QAOA, which in turn requires only $48$ CNOT gates with 3 steps. 

The downside of QAOA, however, is the minimization process, which requires an accurate determination of the optimal parameters. Although classical machines can simulate the process with exact exponentiation for small systems, these become problematic for larger systems. Moreover, it is uncertain whether the minimization process can be performed efficiently in the hybrid computation scenario with NISQ machines. Furthermore, the results reported in the table are the best results out of five runs of the optimizer, as the stochasticity of the simulated annealing method leads to slightly different results. In principle, the variational system of Eq.~\ref{eq:qaoa} can have several local minima, and any algorithm could get stuck in one of them instead of the global minimum which corresponds to the true ground state. However, the qualitative conclusions drawn from comparing QAOA with ASP remain unchanged, as the former is orders of magnitude better for comparable algorithm lengths. In summary, the results indicate that once a set of optimal coefficients $\vec{\gamma}^*$, and $\vec{\beta}^*$ is determined, QAOA is more effective than ASP in preparing the ground state.

\subsection{Blocked QAOA}
\label{sec:blocked-qaoa}
To reduce the number of CNOT gates per qubit, an option is to employ custom optimized 2-qubit gates, which can be tailored for any unitary operation on two qubits, as proposed in~\cite{vatan2004optimal, shende2004minimal}. However, because of nonlocality of the Hamiltonian in Eq.~(\ref{eq:hamiltonian}) owing to the presence of the $H_{ZZ}$ term, this task is not straightforward. Nevertheless, the QAOA algorithm can still be used since it relies on a relatively general ansatz. To handle the nonlocal term, we introduce a modified Hamiltonian, denoted as ``blocked'' or $H_B$, which retains only the diagonal and nearest-neighbor terms of the full Hamiltonian. Thus, defining $H_B = H_\pm + H_Z + H_{ZZ}^l$ this can be represented by a single 2-qubit gate, 
where $H_{ZZ}^l$ is the local part of $H_{ZZ}$. We then modify the QAOA ansatz as follows:
\begin{align}
    \label{eq:ansatz-block}
    \ket{\psi_M( \vec{\gamma}, \vec{\beta})} &= e^{-i\beta_{M} H_0}e^{-i\gamma_{M}H} \\ \nonumber
     & \qquad \times \left(\prod_{k=1}^{M-1}e^{-i\beta_{M-k} H_0}e^{-i\gamma_{M-k} H_B}\right)\ket{\psi_0}.
\end{align}
In the above equation the first $M-1$ unitary applications involve $H_B$, and only one application of the full Hamiltonian is applied on the final step. The aim is to encode the nearest-neighbor interactions using the blocked Hamiltonian, while the last step should adjust for nonlocal effects. This approach can be implemented only when $H_{ZZ}$, which comprises nonlocal terms, is not the dominant term of the Hamiltonian. Thus, we are restricted to cases where $J$ is not large. The results for the blocked approach in the $N=4,8$ case are presented in table~\ref{tab:qaoa-block}.

\begin{table}[!hbtp]\footnotesize
    \centering
    \setlength\tabcolsep{5pt} 
    \begin{tabular}{ccccccc}
        \makecell{\\Method} & \makecell{\\$N$} & \makecell{\\$(\theta,m)$}   & \makecell{\\$M$} & \makecell{CNOT/\\qubit}  & \makecell{Rel.\\Err. $E_0$} & \makecell{GS\\Overlap} \\\hline 
        QAOA   & $4$ & $(0,0)$        & $2$ & $27$              & $0.0026$        & $0.9853$   \\
        QAOA   & $4$ & $(0,0)$        & $3$ & $39$              & $0.0022$        & $0.9922$   \\
        QAOA   & $4$ & $(\pi/4,0.1)$  & $2$ & $30$              & $0.0019$        & $0.9887$   \\
        QAOA   & $4$ & $(\pi/4,0.1)$  & $3$ & $45$              & $0.0015$        & $0.9941$   \\\hline
        QAOA   & $8$ & $(0,0)$        & $2$ & $38.5$            & $0.0028$        & $0.9553$   \\
        QAOA   & $8$ & $(0,0)$        & $3$ & $52.5$            & $0.0024$        & $0.9632$   \\
        QAOA   & $8$ & $(\pi/4,0.1)$  & $2$ & $38.5$            & $0.0023$        & $0.9763$   \\
        QAOA   & $8$ & $(\pi/4,0.1)$  & $3$ & $52.5$            & $0.0034$        & $0.9711$   \\
    \end{tabular}
    \caption{\footnotesize Blocked QAOA results for $N=4$. Here $N$, $\theta$ and $m$ are the parameters of the simulation, $M$ is the number of QAOA steps. The remaining columns give the CNOT gates per qubit, the relative error of the ground state energy measurement, and the ground state overlap.}~\label{tab:qaoa-block}
\end{table}

The decrease in the number of CNOT gates is more noticeable for larger values of $N$. An exact expression for this type of blocking would be: $8M(N-1) + (N-1)(N-2)$. One can note that there is no significant difference between the results with the full and blocked QAOA ansatz. This implies that the non-local part of the Hamiltonian can be encoded efficiently just in the last step.
Consequently, a possible advantage of the blocking procedure is the potential to 
scale the system size while maintaining the optimal parameters $\vec{\gamma}$ or 
$\vec{\beta}$ fixed as $N$ changes. 
This approach is not perfect as the non-local part of the Hamiltonian 
changes with increasing system size. 
However, it produces good ans\"atze for 
larger systems without requiring costly optimization procedures. 
Table~\ref{table:qaoa_scaling} illustrates the results for different 
system sizes using the same QAOA parameters obtained through simulated 
annealing on the $N=4$ system for the $(\theta,m)=(0,0)$ case with three steps, 
two of which are blocked while the last one is executed with the full Hamiltonian.

\begin{table}[!htbp]\footnotesize
    \centering
    \setlength\tabcolsep{5pt} 
    \begin{tabular}{ccccc}
        \makecell{\\$N $} & \makecell{CNOT/\\qubit} & \makecell{Rel.\\Err. $E_0$} & \makecell{GS\\Overlap} \\\hline
        $4*$ & $39$            & $0.0003$        & $0.9995$    \\
        $6 $ & $46.6$          & $0.0022$        & $0.9839$    \\
        $8 $ & $52.5$          & $0.0021$        & $0.9722$    \\
        $10$ & $57.6$          & $0.0022$        & $0.9599$    \\
        $12$ & $62.3$          & $0.0020$        & $0.9479$    \\
    \end{tabular}
    \caption{\footnotesize QAOA blocked with $M=3$ steps. The results for the $N=4$ are obtained after parameter optimization; the results for $N \in \{6,8,10,12\}$ have been computed using the same optimal parameters for $N=4$. }~\label{table:qaoa_scaling}
\end{table}

The data presented indicates that blocked QAOA can serve as a very cheap starting point for subsequent optimizations or alternative state-preparation algorithms, particularly when dealing with large values of $N$.

\section{Rodeo Algorithm}
\label{sec:rodeo}

The recently proposed rodeo algorithm (RA)~\cite{Choi:2020pdg, Qian:2021arx, BeeLindgren2022} is a promising method that uses a stochastic cosine filter to isolate eigenstates of a given Hamiltonian. It can be used to extract the energy spectrum of a Hamiltonian, and it can also be used as a state-preparation algorithm. One of its biggest advantages is that it allows one to prepare a system in any eigenstate of the Hamiltonian, not just the ground state.

The rodeo algorithm uses one or more ancilla qubits to control the time evolution of the initial object state $\ket{\psi_I}$ by the object Hamiltonian $H$. In the general case, the $M$-cycle RA uses a set of $M$ ancilla qubits. However, for quantum computers that allow mid-circuit measurements, a single ancilla qubit may be used repeatedly~\cite{Qian:2021arx}. The algorithm starts with all ancilla qubits in the same state e.g. $\ket{1}$. A Hadamard gate is applied to fully mix each ancilla qubit. Then for the $m$th control/ancilla qubit, a controlled time evolution $e^{-i\hat{H}t_m}$ is applied to the object system, followed by a phase rotation $P(Et_m)$ applied to the ancilla qubit. In the end, a Hadamard gate is applied to each ancilla qubit, and it is measured.

\begin{figure}[!hbpt]
    \centering
    \includegraphics[width=1\linewidth]{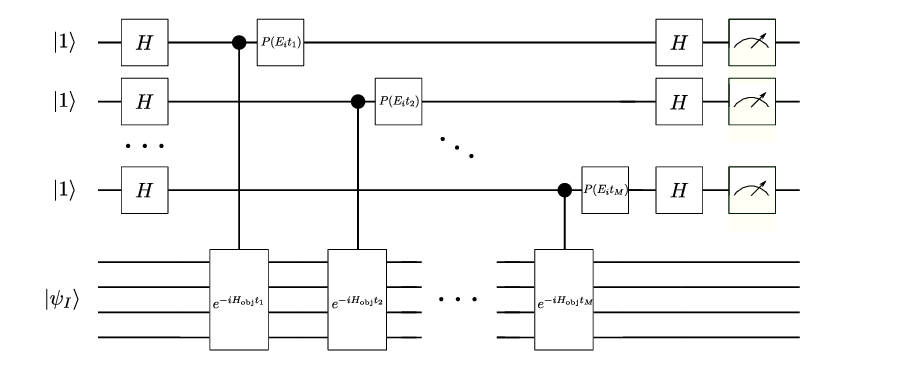}
    \caption{\footnotesize Gate representation of the rodeo algorithm.}
    \label{fig:rodeo-algorithm}
\end{figure}

Let us consider a single rodeo cycle. Starting from the initial state $\ket{1}\ket{\psi_I}$, after performing one cycle of the RA and inserting a complete set of energy eigenstates, the system is in the state
\begin{align}
    \begin{split}
        &          \frac12 \sum_j \braket{E_j|\psi_I } \left(1- e^{-i(E_j- E)t_1} \right) \ket{0} \ket{E_j} \\
        & \qquad + \frac12 \sum_j \braket{E_j|\psi_I } \left(1+ e^{-i(E_j- E)t_1} \right) \ket{1} \ket{E_j}
    \end{split}
\end{align}
The probability, as a function of $E$, of measuring the ancilla qubit in the original state $\ket{1}$ is
\begin{equation}
    P_{\ket{1}}(E) = \sum_j \left|\braket{E_j|\psi_I }\right|^2 \cos^2\left([E-E_j]\frac{t_1}{2}\right) .
\end{equation}
Thus, if we take random values of the evolution time $t_m$, we have a cosine filter for the energy, which can be tuned to exponentially suppress eigenstates outside an energy range. For large $M$, the spectral weight for any eigenstate with $E_j \neq E$ is suppressed by a factor of $1/4^M$.

\begin{figure}[!hbpt]
    \centering
    \includegraphics[width=1\linewidth]{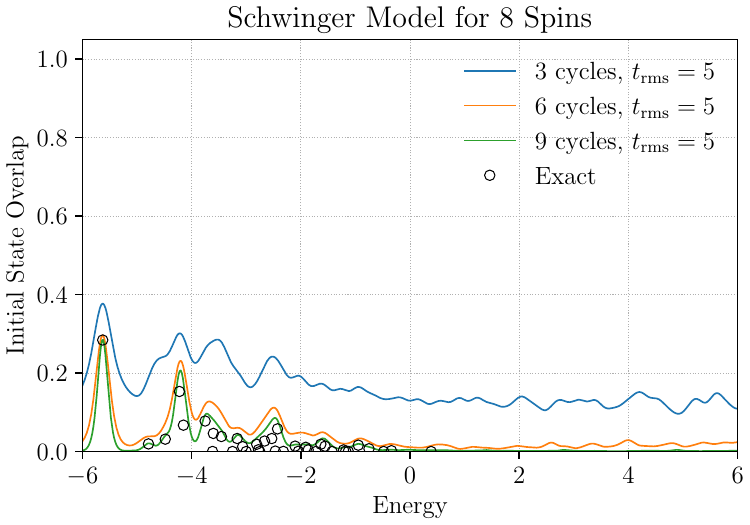}
    \caption{\footnotesize Spectral overlap factors for the $\ket{10101010}$ initial state from the rodeo algorithm.}
    \label{fig:rodeo_spectrum}
\end{figure}

The RA can be used to extract the spectrum of the Hamiltonian $H$. If we label the eigenstates of $H$ as $\ket{E_j}$, we can define the initial-state spectral overlap function as $S(E) = |\braket{E_j|\psi_I}|^2$ for $E=E_j$ and $S(E)=0$ for $E\neq E_j$. In practice, one runs the RA at some fixed energy $E$ and random value $t_m$ for each rodeo cycle. This is repeated with a set of Gaussian random values for the times $t_m$, and the results are averaged over to get the value of $S$ at $E$. The function $S(E)$ is constructed by repeating this procedure for a range of values $E$. An example for the Schwinger model is shown in Fig.~\ref{fig:rodeo_spectrum}.

The RA however has limitations for NISQ machines. First, it requires an additional number of qubits, the ancilla, that are not available for the simulation after the state is prepared. Furthermore, the stochastic values for the controlled time evolution have to be chosen from a distribution with a certain width $\sigma$. This unphysical parameter controls the width of the cosine filters that are effectively applied. If a too small value is chosen the filtering effect is not sufficient to isolate the different peak corresponding to distinct eigenvalues. But choosing a too large value necessitates simulating large times with a time evolution operator, increasing the necessary Trotter steps.

\subsection{Rodeo Algorithm for State Preparation}
\label{subec:rodeo_state_prep}

The RA suppresses eigenstates whose eigenvalues differ from $E$ and thereby effectively amplifies any eigenstate whose energy is close to $E$. This means the RA can be used for state preparation.

If the exact energy $E_k$ of the target eigenstate $\ket{E_k}$ is known, then one simply performs some number $M$ of rodeo cycles at the energy $E=E_k$. In the end, if all ancilla qubits are measured to be in the state $\ket{1}$, then the object state is $\ket{E_k}$ with a probability which ranges from the initial state overlap probability $|\braket{E_k|\psi_I}|^2$ when $M=0$ to 1 in the limit of large $M$. If any ancilla qubits are measured to be in the state $\ket{0}$, then the state is discarded and one restarts the algorithm. There are two probabilities to keep in mind: the probability of measuring all ancilla qubits to be in the state $\ket{1}$, and the conditional probability that the object state is $\ket{E_k}$ given that all ancillas are in the state $\ket{1}$. In the limit of large $M$, the first probability is equal to the initial state overlap probability, and the second is equal to one. The cost, in terms of gates and circuit depth, of the RA for state preparation depends on several factors including the initial state overlap with the desired eigenstate, and the number of rodeo cycles $M$. The controlled time evolution can be realized through Toffoli gates, which extend CNOT gates to have 2 control qubits. These in turn can be decomposed in a series of 6 CNOT gates and single qubit rotations. In general, the cost of the RA for our Schwinger Hamiltonian, with first-order trotterization, is then: $6M[4(N-1) + (N-1)(N-2)]$. 

\begin{figure}[!hbpt]
    \centering
    \includegraphics[width=1\linewidth]{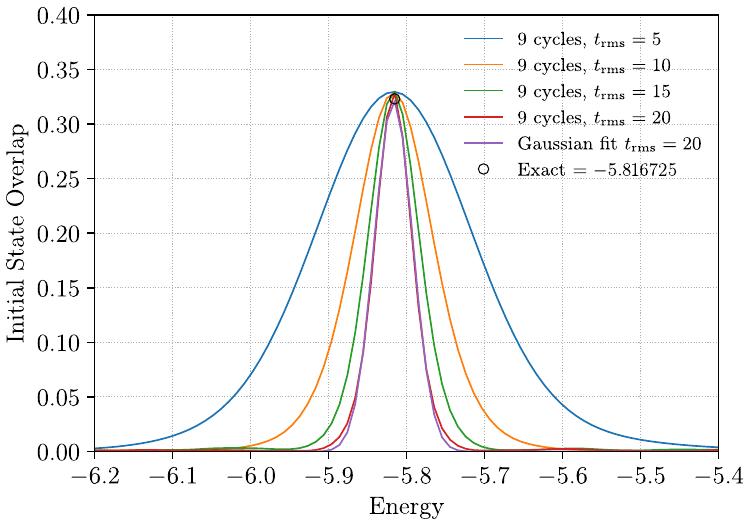}
    \caption{\footnotesize Gaussian fit to the data of the ground state peak from Fig.~\ref{fig:rodeo_spectrum}.}
    \label{fig:rodeo_peak}
\end{figure}

If the energy $E_k$ of the target eigenstate $\ket{E_k}$ is not known, then the RA can be used to scan for the precise energy by trying a range of energies $E$ as in Fig.~\ref{fig:rodeo_spectrum}. Once the relevant peak is isolated in the spectral overlap function, its precise location can be extracted using a Gaussian fit as shown in Fig.~\ref{fig:rodeo_peak}. The algorithm is then repeated at this energy.

\subsection{Preconditioning the Rodeo Algorithm with QAOA}
The RA efficiency greatly improves when the overlap of the initial state with the desired state is large. In particular, the number of required cycles in the RA and their total time evolution length, and hence the overall length of the algorithm, can be reduced. One can then use a state coming from the blocked QAOA presented in Section~\ref{sec:blocked-qaoa}, which is cheap to prepare, and use it as initial state for the RA.\

\begin{figure}[!hbpt]
    \centering
    \includegraphics[width=1\linewidth]{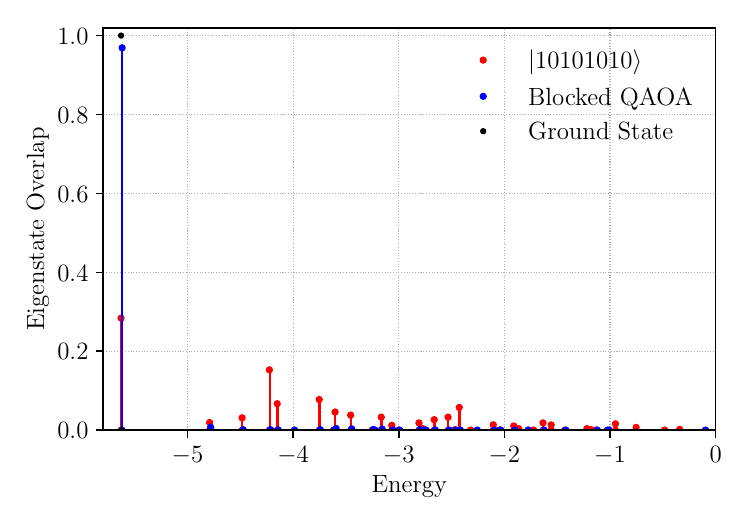}
    \caption{\footnotesize Comparison of the eigenstate overlaps between the QAOA ansatz and the simple alternating spin initial state.}
    \label{fig:initial-comparison}
\end{figure}

The procedure is as follows: first, a classical optimization of the blocked QAOA model is made on a small system, in our case we used the $N=4$ Schwinger model, the same of Table~\ref{tab:qaoa-block}. Secondly, the state coming from the QAOA ansatz for a larger system is prepared. As seen in the table, for the $N=8$ the overlap $\approx 96\%$, which is considerably better as an initial guess when compared to the alternating chain of spins up and down, see Fig.~\ref{fig:initial-comparison}.
Finally, one can perform just 3 cycles of the RA with a small time evolution step, in practice, we restrict the random times $t_m$ to have a root-mean-square value of $t_\mathrm{rms=1,2}$, and perform a scan of the energies close to the ground state and use the fitting 
procedure outlined in Sec.~\ref{subec:rodeo_state_prep}. Keeping the time short is crucial to reduce the number of Trotter steps required to perform the  controlled time evolution. In particular, since the values of $t_m$ are stochastic, we fix a value for the time steps of $\delta t = 0.25$, with the last step being shorter depending on the exact value of $t_m$. 

\begin{figure}[!hbpt]
    \centering
    \includegraphics[width=1\linewidth]{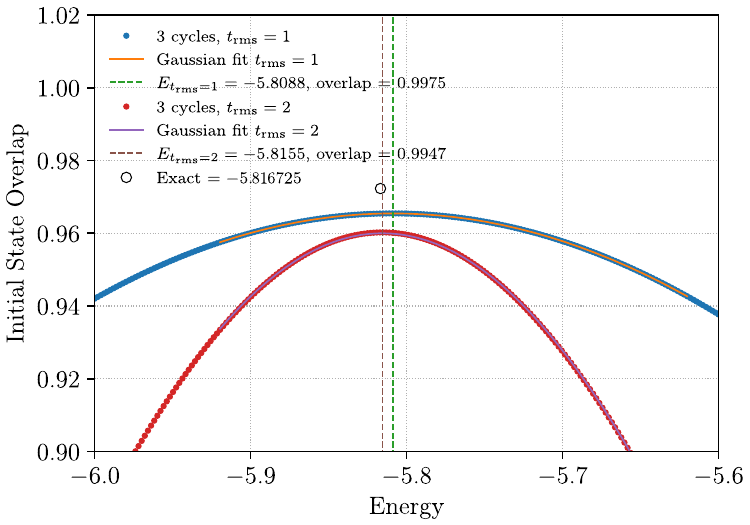}
    \caption{\footnotesize Gaussian fit to the data of the ground state peak from located starting from the blocked QAOA ansatz for $N=8$ and $t_{rms}=1,2$.}
    \label{fig:rodeo_peak_qaoa}
\end{figure}

As seen in Fig.~\ref{fig:rodeo_peak_qaoa}, the ground state energy can be determined 
with a good degree of accuracy with very few cycles and a short time evolution.

The advantage of combining the blocked QAOA and the RA is that one can perform the classical-hybrid optimization for large systems in QAOA and correct for the error coming from the mistuned parameters using the RA, which in turn benefits in efficiency from the improved initial ansatz. 
The total algorithm for the blocked QAOA preconditioned RA for $N=8$ cannot be estimated exactly as the length depends on the random choices for the various $t_m$. However, with $t_{rms}=1$ and $\delta t = 0.25$ and $3$ cycles one can expect that on average the CNOT count will be on the order of 12 time evolution steps plus the initial QAOA preconditioning. Considering the decomposition of the Toffoli gates, we find that on average the total count of CNOT gates per qubit is 661, which corresponds to 75 adiabatic steps, while reproducing the ground state with 99.5\% accuracy.
It should be noted that this procedure retains the requirement to check the values of the ancilla qubits of the RA to determine whether the state prepared is valid or not.

\section{Summary and Conclusions}
\label{sec:conclusion}

In this paper, we explore efficient quantum state preparation algorithms for the Schwinger model with a theta term, focusing on adiabatic state preparation (ASP), the quantum approximate optimization algorithm (QAOA), and the rodeo algorithm (RA). We have analyzed these algorithms based on their efficiency, scalability, and quantum resource requirements, with particular focus given to the length of the algorithms as measured in terms of the total number of two-qubit gates, as this is a useful metric for their performance on noisy intermediate-scale quantum (NISQ) machines.

While ASP has its strength in its simplicity and interpretability, it requires many time-evolution steps to reach good accuracy as it is bounded by both numerical discretization in the size of the steps and by the adiabatic theorem for the total time of the evolution. We have found that choosing different discretizations for the size of the time-evolution steps can lead to improved accuracy, but that is not a significant improvement as the overall scaling is the same.

The QAOA produces a short algorithm with high precision, but it requires a classical optimization step which can be extremely costly and doesn't scale with the system size. We have implemented a blocked ansatz for the QAOA that enables scaling the system by classically optimizing the parameters for a small system and reusing them for larger ones, though the precision decreases with the scaling.

The RA excels in preparing any eigenstate, not just the ground state. Its success depends on the overlap between the initial and target states, which can be optimized using other algorithms.
We find that combining blocked QAOA with the RA provides the best results. Blocked QAOA efficiently prepares a high-quality initial state, which enhances the performance of the RA by reducing the number of required cycles and improving the accuracy of the state preparation. For the 8-site system, we find that our procedure returns states that have 99.5\% overlap with the true ground state with the equivalent CNOT gate counts of just 75 adiabatic steps.

The combination of blocked QAOA and the RA presents a scalable and resource-efficient method for state preparation in quantum simulations of the Schwinger model. This hybrid approach minimizes gate counts and classical optimization complexity, making it well-suited for larger systems and NISQ-era quantum devices. The results demonstrate promising directions for future research in quantum algorithms for quantum field theory simulations, paving the way for applications in more complex models.

\section*{Acknowledgments}

This work was supported by the Department of Energy grant DE-SC0021152. Additionally, D.~L. was supported by Department of Energy grant DE-SC0023658 and National Science Foundation grant PHY-2310620, and A.~B. and L.~H. were supported by Department of Energy grant DE-SC0019139. G.~P. acknowledges funding by the Deutsche Forschungsgemeinschaft (DFG, German Research Foundation) - project number 460248186 (PUNCH4NFDI).
A.S. acknowledges funding support from Deutsche Forschungsgemeinschaft 
(DFG, German Research Foundation) through grant 513989149,  
under the National Science Foundation grant PHY- 2209185 and from the Department of Energy Topical Collaboration “Nuclear Theory for New Physics”, award No. DE-SC0023663


\bibliography{schwinger_theta}

\end{document}